\documentclass[%
reprint,
showpacs,preprintnumbers,
 amsmath,amssymb,
 prl,
]{revtex4-1}

\usepackage{graphicx}
\usepackage{dcolumn}
\usepackage{bm}

\begin{document}


\title{Delocalization of Nonlinear Optical Responses in Plasmonic Nanoantennas}

\author{Sviatlana Viarbitskaya}
 \email{sviatlana.viarbitskaya@u-bourgogne.fr}
\author{Olivier Demichel}%
\author{Benoit Cluzel}%
\author{G\'erard Colas des Francs}%
\author{Alexandre Bouhelier}%
 \email{alexandre.bouhelier@u-bourgogne.fr}
\affiliation{%
Laboratoire Interdisciplinaire Carnot de Bourgogne CNRS-UMR 6303, Universit\'e de Bourgogne, 21078 Dijon, France.
}%

\date{\today}

\begin{abstract}
Remote excitation and emission of two-photon luminescence and second-harmonic generation are observed in micrometer long gold rod optical antennas upon local illumination with a tightly focused near-infrared femtosecond laser beam. We show that the nonlinear radiations can be emitted from the entire antenna and the measured far-field angular patterns bear the information regarding the nature and origins of the respective nonlinear processes. We demonstrate that the nonlinear responses are transported by the propagating surface plasmon at excitation frequency, enabling thereby polariton-mediated tailoring and design of nonlinear responses.
\end{abstract}

\maketitle

Optical antennas are pervasive devices to control spatial distribution of light on sub-diffraction length scales~\cite{Schuller10, novotny11NP, OptAntBook13}. Concurrently it is realized that field enhancing properties of underlying surface plasmon (SP) resonances may foster much needed nonlinear behaviors to improve nanoscale light management \cite{Kauranen:2012aa,NovotnyNLbook,NLObook}. A nonlinear optical antenna combines the functionalities of linear devices (extreme light concentration, tailoring of spatial and phase distributions, directivity of emission, etc) with the benefits of nonlinear optical effects, such as frequency conversion~\cite{kimNature08}, mixing~\cite{Stolz2014}, ultrafast switching, modulation~\cite{Muskens11,Wurtz11} and self-action effects~\cite{DeLeon:14,Baron:15}, to name just a few. Nonlinear responses are notably intricate and a comprehensive theory of nonlinear plasmonics is still underway. To date the most extensively elaborated are harmonic generation, four-wave mixing and multi-photon luminescence at the single optical nanoantenna level~\cite{Dadap04,orrit05NL,bouhelier05PRL,Bachelier08,sipe09,danckwerts07,Ginzburg10,Cai11,metzger14,ginsburg14}. It has been repeatedly demonstrated that nonlinear spectral and intensity responses in plasmonic antennas are largely determined by the localized SP resonances at the frequencies of the driving optical fields \cite{hecht05sciences,NovotnyNLbook,Novotny_11,Knittel:15}. In spatially extended plasmonic objects, point-and-probe nonlinear scanning microscopy revealed the importance of the supported SP modal landscape~\cite{quidant08, Berthelot:12,valev:12, viarb13nm, Demichel:14}, suggesting that the nonlinear responses may bear the signatures of the SP mode spatial extension. It is this effect that we address in this letter. 

To this aim, we discuss two nonlinear processes - two-photon luminescence (TPL) and second-harmonic generation (SHG) - from gold rod optical antennas upon local illumination with a tightly focused femtosecond near-infrared laser beam. We show that in this type of structures, nonlinear confocal TPL and SHG mappings are ineffective when it comes to discerning differences between these two processes. The variations between incoherent TPL and coherent SHG are unambiguously revealed in spectrally filtered angular distributions measured in Fourier and image planes. Importantly, we demonstrate that nonlinear conversions of the incident electromagnetic energy are not restricted to the excitation area but are spatially delocalized along the entire structure. We argue that nonlinear optical transport is mediated by a propagating SP at the excitation frequency despite the associated high losses. We substantiate this hypothesis by modelling far-field SHG signatures as originating from a line of non-interacting dipolar sources oscillating at the SHG frequency, whose amplitudes and phases are determined by the damped SP polariton at the fundamental frequency, developing in the one-dimensional cavity-like antenna. Our findings demonstrate new degrees of freedom for design of SP mediated coherent and incoherent nonlinear processes.

Optical gap antennas are fabricated by electron-beam lithography and lift-off technique on a glass substrate. Each antenna consists of two identical nanowires separated by a gap. The dimensions of individual wires are 110 nm in width and 50 nm in height. The length ${\it L}$ of each arm and the gap separation ${\it g}$ are systematically varied from 300 nm$<L<$4000 nm, and from 0 nm$<{\it g}<$150 nm, with a minimum gap size of ca. 20 nm, as measured with scanning electron microscope (SEM). Optical excitation and collection are performed using an inverted microscope. A 180 fs Titanium: Sapphire laser tuned at a fundamental wavelength of $\lambda_0=$810 nm is focused on the antennas in a diffraction-limited spot by a high numerical aperture objective (oil immersion, NA=1.49). The average laser power at the sample is 3.5 mW. The incident beam is linearly polarized along the antenna. Nonlinear signals are collected by the same objective followed by a dichroic beam splitter, which separates the useful spectral range (375-700 nm) from the back-scattered fundamental beam. Simultaneous TPL and SHG confocal maps are collected by two avalanche photodiodes in the absence of a spatial filter, allowing detection of signal emitted from the entire structure. A 10 nm narrow bandpass filter centered at 405 nm is used for SHG detection. Fourier and direct planes imaging are recorded by separate cameras and relay lenses appropriately placed in their respective conjugate planes~\citep{SongJOVE}.  

\begin{figure}[!t]
\includegraphics{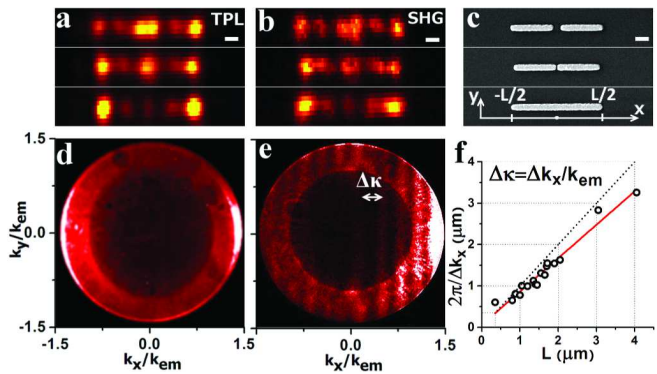}
\caption{\label{fig:1} (a)-(c) TPL, SHG confocal maps and SEM micrographs of three gold rod antennas. The arm lengths of the coupled antennas are 650 nm, the gaps are $g$=130 and 50 nm. (d) and (e) Angular distributions of TPL and SHG signal from a single nanowire (ca.1500 nm long, the bottom antenna in (a)-(c)). Scale bars are 200 nm. The laser is focused at the left extremity. The images are normalized in units of $k_{em}=2\pi/\lambda_{\rm em}$, where $\lambda_{\rm em}$ is the emission wavelength of the nonlinear process. The inter-fringe distance $\Delta \kappa$ is denoted with the white double-headed arrow. (f) Linear dependence of $2\pi/ \Delta k_{x}$ on nanowire length $L$ (solid line is the linear least square fit) and $2\pi/ \Delta k_{x}=L$ (dotted line). }
\end{figure}

TPL is an incoherent nonlinear optical process defined by the material's electronic band structure~\cite{boyd86,BiagoniNL12,Tigran13} and underlying plasmonic modes~\cite{bouhelier05PRL,viarb13nm,verellen15,Knittel:15}. In conjunction with confocal mapping technique, it probes local electric field intensity in plasmonic objects~\cite{quidant08, hechtNL10}. Similarly, SHG point-and-probe mapping provides some additional information about material due to the intrinsic dependency of SHG on structural symmetries~\cite{kauranen07,Berthelot:12,Czaplicki14,valev:12}. Having applied both methods to the lithographic rod antennas, we find that, apart from the difference in the nonlinear yield (typically TPL/SHG ratio is at least $10^3$), confocal nonlinear maps are practically identical as illustrated in Fig.~\ref{fig:1}(a) and (b). This was true for all studied optical gap antennas, regardless of the length or the gap size. Because of this striking similarity, we rule out any significant structural symmetry-dependent SHG component, which could render SHG maps to be somewhat different from the TPL ones~\cite{NovotnyNLbook}. It follows that the spatial variations of both SHG and TPL signals must be accounted for by the same sensitivity to the local plasmonic modal distribution~\cite{viarb13nm,deCeglia:15}.

While nonlinear confocal mapping fails at distinguishing between these fundamentally different nonlinear processes, Fourier plane imaging unveils the information hidden in the pixels of the nonlinear confocal maps. In Fig.~\ref{fig:1}(d) and (e), we compare Fourier plane images representing the projected angular distributions of the TPL and SHG emissions from a single rod nanoantenna [bottom nanowire in Fig.~\ref{fig:1}(a)-(c)]. When the laser is focused on the left extremity, the TPL Fourier distribution [Fig.~\ref{fig:1}(d)] features a pattern with two maxima aligned along the antenna's $x$ axis coinciding with the excitation polarization direction at variance with a single x-oriented dipole~\cite{Hartmann13}. Vastly different is the SHG angular image shown in Fig.~\ref{fig:1}(e) displaying distinct interference fringes reflecting the coherent nature of SHG. Excitation at the left extremity of the antennas systematically produces fringe patterns with an intensity increasing towards positive $k_{x} /k_{\rm em}$. Such fringe pattern was already observed in the linear regime for nanowires decorated with emitters~\cite{Hartmann13}. SHG Fourier images show a strict dependence on the rod length. Figure~\ref{fig:1}(f) shows the reciprocal dependence of the fringe period $\Delta k_{x}$ on the antenna length. As TPL is intrinsically an incoherent process, we do not observe interferences in Fourier plane even when a narrow portion of the broad TPL spectrum was spectrally-selected. Neither do the TPL Fourier images depend on the antenna's length and patterns similar to that in Fig.~\ref{fig:1}(d) are persistently obtained.

\begin{figure}[!t]
\includegraphics{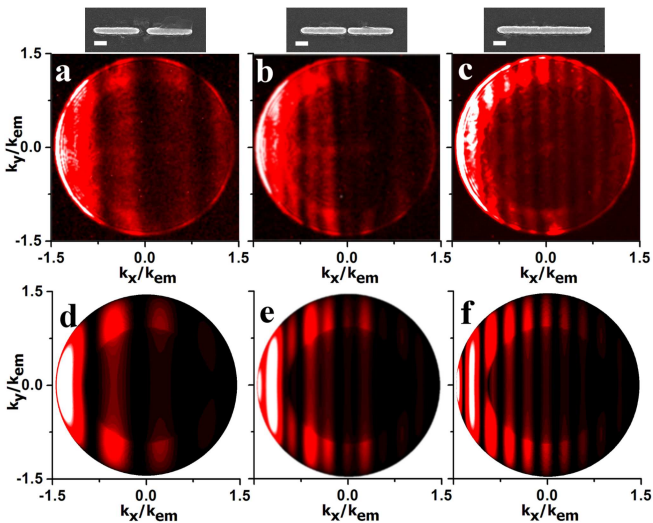}
\caption{\label{fig:2} (a)-(c) Evolution of Fourier plane patterns with the gap size: 140 nm, 40 nm, 0 nm. Antennas arm length is ca. 830 nm, excitation at the right-hand extremity. Simulated Fourier planes: (d)  $L_{\rm sim}$=500 nm, (e) far-field beating between $L_{\rm sim}$=500 nm and 1400 nm, (f) $L_{\rm sim}$= 1400 nm. The effective refractive index at $\omega$ is  $n_{\rm eff}$ =1.73, the propagation length is $L_{\rm sp}=1/2k_{\rm sp}''$ = 2100 nm.}
\end{figure}

Fourier plane imaging of SHG response of coupled optical gap antenna shows sensitivity to the gap size. Figure~\ref{fig:2}(a)-(c) illustrate a set of Fourier planes obtained from three antennas having identical arm lengths (ca. 830 nm) and gaps of 140 nm, 40 nm and at contact, correspondingly. The excitation is located at their respective right extremities. For maximally decoupled arms [$g$=140 nm, Fig.~\ref{fig:2}(a)], we observe four wide fringes, which, as the gap size decreases, start to split [Fig.~\ref{fig:2}(b)], forming a pattern of a different symmetry. The splitting becomes more pronounced as the gap decreases to its minimum. A Fourier pattern corresponding to the case of touching arms [Fig.~\ref{fig:2}(c)] contains a double number of fringes as compared to the case of maximally separated arms. We attribute such splitting/pairing of the fringes to the onset of coupling between the antenna's arms.

The sensitivity of SHG Fourier planes to the antenna's length and the gap size indicates that the nonlinear response is not simply generated locally by the focused laser beam. At this point, we note that the excitation of the nanorod extremity by a focused laser beam is a prevalent technique to launch a propagating SP in the structure~\cite{dickson00,SongACS11,Xu12review}. In order to account for the SHG Fourier planes' length dependence, in the following we assume that the propagating SP is creating an enhanced electric field along the antenna strong enough to allow remote nonlinear optical interactions. We construct a model of the delocalized SHG emission from a single gold antenna of length $L$ by considering propagation of the SP in one-dimensional cavity (ODC) of length $L_{\rm sim}$~\cite{Taminiau11}. Upon point dipole excitation at the cavity's left extremity ($x=-L_{\rm sim}/2$), which we equate here with the focused laser beam excitation, the plasmon associated electric field along the antenna $E_{sp}^{\omega}(x)$ can be written as~\cite{Taminiau11}:
\begin{multline}
E_{\rm sp}^{\omega}(x)= E_0e^{ik_{\rm sp}L_{\rm sim}/2}(r+1)\frac{(e^{ik_{\rm sp}x}-re^{ik_{\rm sp}(L_{\rm sim}-x)})}{1-r^{2}e^{i2k_{\rm sp}L_{\rm sim}}},\label{final}
\end{multline}
where $E_0$ is the excitation field amplitude, $r$ is the reflection coefficient. The values of complex SP wave vector $k_{\rm sp}=k_{\rm sp}'+ik_{\rm sp}''$ are extracted from finite element simulations of an infinitely long nanowire with the cross section of the nanowires used in the experiment. The nonlinear emission is phenomenologically modelled by a large number of identical non-interacting effective dipoles $p^{2\omega}$, oscillating at the SHG angular frequency. The dipoles are aligned and oriented along the antenna $x$ axis on the glass/air interface. The amplitudes and phases of the individual oscillators are position-dependent and determined by the ODC scalar plasmon wave $ E_{\rm sp}^{\omega}(x)$, according to $p^{2\omega}(x)=\beta^{(2)}[E_{\rm sp}^{\omega}(x)]^{2}$, where $\beta^{(2)}$ is a 1D counterpart of the nonlinear polarizability of the individual effective dipoles \cite{HEINZ1991}. This treatment of SHG is inspired by the earlier works establishing the link between the dominant SHG origin with the local symmetry breaking at the surface~\cite{Bachelier08,sipe09}. Note that in our case the interband structure of gold inhibits plasmon modes at the SHG frequency due to strong absorption and prevents tailoring of plasmonic modes at the harmonic frequency~\cite{Thyagarajan:12}.

We employ the known expressions of the far-field Green functions to calculate the coherent sum of electric fields emitted by each SHG dipole source, collected by the objective~\cite{lukasbook}. The results of simulations for the line of $p^{2\omega}$ dipoles with the total length of $L_{\rm sim}$=1400 nm and 500 nm are shown in Fig.~\ref{fig:2}(d) and (f). In the presence of a gap, we calculate the Fourier plane as a coherent beating between far-field signals emitted from two nanowires of $L_{\rm sim}$=500 nm and $L_{\rm sim}$=1400 nm. The result of the calculation is shown in Fig~\ref{fig:2}(e), which reproduces well the experimental fringe splitting when short (S) and long (L) lengths contribute to the final signal with equal weights $w_{S}=w_{L}$. We find a good match between experimental and simulated data when $L_{\rm sim}$ is close to the effective nanowire length $2\pi/ \Delta k_{x}$ [Fig.~\ref{fig:1}(f)]. We think that the systematic difference between the physical length $L$ and $L_{\rm sim}$ could be due to intrinsic oversimplification of the real electric field in the nanowires inherent to the OCD model. Among other factors is the omission of other possible mechanisms of SHG. The latter will be particularly interesting to investigate in the light of the recent discussions on ponderomotive force, Kerr-like and heat induced nonlinearities in metals~\cite{ginzburg15, Marini:13}.

To accurately fit the experimental SHG Fourier images, the coefficient $r$ in Eq.~\ref{final} is set to zero. This implies that there is no contra-propagating field components in the cavity, which can be understood by large Ohmic losses and efficient end-face scattering~\cite{SongACS11}. Interestingly, our simulations predict that if even a small portion of the electromagnetic excitation could scatter back into the cavity, a drastic change would be observed in the SHG Fourier pattern, which is not the case here. In the absence of a back-reflected SP, the ODC model prediction of the scalar field $E_{\rm sp}^{\omega}(x)$ coincides simply with a damped wave $e^{ik_{\rm sp}(x+L_{\rm sim}/2)}$, as can be easily seen from Eq.~\ref{final}. In this case, the Fourier planes are merely Fourier transforms of the $e^{2ik_{\rm sp}(x+L_{\rm sim}/2)}$ function, and the observed fringes are so-called Gibbs oscillation, resulting from the finite length of waveguide~\cite{Hassan13}. 

\begin{figure}[!t]
\includegraphics{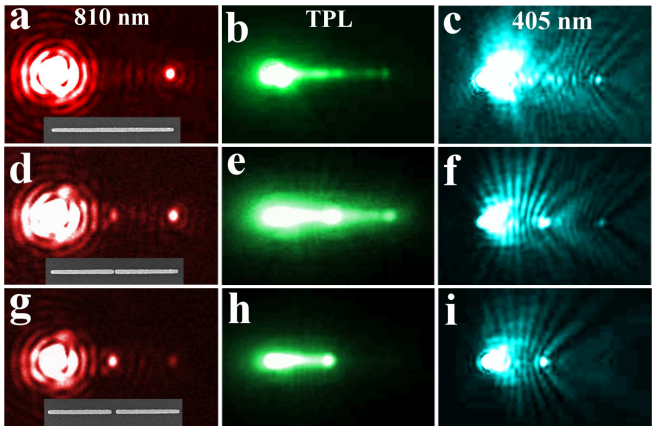}
\caption{\label{fig:3} (a), (b) and (c) are colorized direct image plane micrographs emphasizing the excitation of propagating SP mode in a 1950 nm-long gold rod nanoantennas.  Images are spectrally filtered to record emission at the fundamental, the TPL, and SHG wavelengths, respectively. The nonlinear signals are delocalized along the entire length of the antenna. (d) to (f) and (g) to (i) are obtained from coupled gap antennas with $g$=50 nm, and $g$=180 nm. Insets are SEM images of the corresponding antennas. Excitation is at the extreme left end.}
\end{figure}

Figure~\ref{fig:3} experimentally confirms the assumptions of delocalized nonlinear responses tested in the simulation. Figure~\ref{fig:3}(a) shows an image plane micrograph of a single rod antenna excited locally by the laser beam focused at its left extremity. The image is recorded at the fundamental wavelength $\lambda_0$. The light detected at the distal end unambiguously indicates the excitation of a SP mode in the antenna~\cite{dickson00}. No leaky plasmon mode is expected in our structures due to their small transverse cross section~\cite{SongACS11}.  The corresponding spectrally filtered TPL and SHG image planes are shown in Fig.~\ref{fig:3}(b) and (c), respectively. Aside from the strong local spot at the laser position, it becomes evident that the TPL response is delocalized along the entire rod [Fig.~\ref{fig:3}(b)]. Local scattering at the rod's defects and structural discontinuities are also readily observed. Similar point scattering is observed in the SHG filtered image of Fig.~\ref{fig:3}(c), together with characteristic interference patterns reminiscent of the fringes detected in Fourier planes. While these nonlinear images are somewhat degraded by residual chromatic aberrations from relay lenses, they unambiguously demonstrate the SP mode developing in the antenna at $\lambda_0$ carries enough energy to produce a distributed nonlinear response during its propagation. To further confirm this conclusion, we studied the delocalized responses for coupled gap antenna with $g$=50 nm and $g$=180 nm. The images recorded at $\lambda_0$ indicate a significant near-field coupling for the smallest gap as the plasmon excitation is transferred from one arm to the other [Fig.~\ref{fig:3}(d)]. For the decoupled antenna, scattering at the gap strongly mitigates the transmission to the right arm [Fig.~\ref{fig:3}(g)]. Concomitantly, the spatial extent of the nonlinear processes are gap dependent. In Fig.~\ref{fig:3}(e)-(i), both TPL and SHG emission are observed from the second arm, in line with the fact that the plasmon at $\lambda_0$ is conveyed through the gap. For decoupled antennas, the intensity of the nonlinear emissions emitted along the second arm disappear because most of the energy carried by the plasmon is scattered at the gap [Fig.~\ref{fig:3}(h)-(i)]. 

\begin{figure}[!t]
\includegraphics{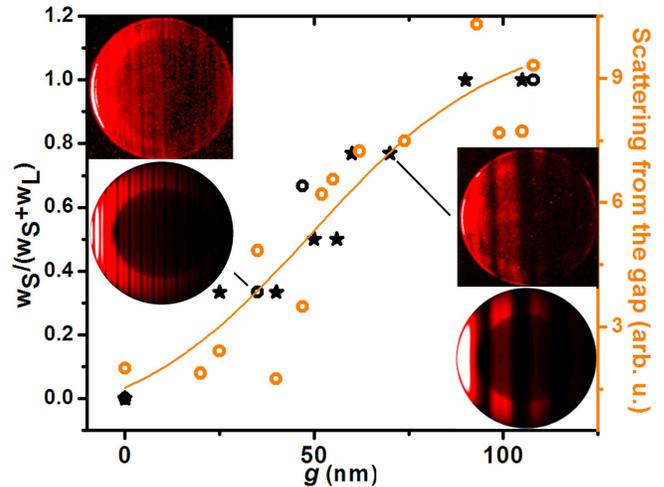}
\caption{\label{fig:4} Correlation between relative linear scattering efficiencies of SP from the gaps of the 1950 nm arm long antenna (orange circles, Fig.~\ref{fig:3}(a),(d) and (g)) and the extracted from the experimental and simulated Fourier plane images values of the mixing parameter $w_{S}/(w_{S}+w_{L})$ (black circles). Black stars are the mixing parameters for the set of ca.~830 nm arm long antennas (Fig.~\ref{fig:2}). The insets are experimental and simulated Fourier planes, showing different degrees of fringe splitting for two selected gap-antennas. The orange curve is added for eye-guiding purposes.}
\end{figure}

Hence, the efficiency of energy transport through the gap upon excitation at the extremity can be indirectly monitored by (i) observing light scattering from the gap in linear regime [Fig.~\ref{fig:3}(d) and (g)], or by (ii) observing the fringe patterns in Fourier planes in nonlinear regime (Fig.~\ref{fig:2}). These two aspects are brought together in Fig.~\ref{fig:4}. The relative scattering efficiency at $\lambda_{0}$ is measured as a function of gap size (orange circles) from antennas shown in Fig.~\ref{fig:3}, where maximal coupling corresponds to the minimal scattering ($g$=0). Then, we plot the quantity $w_{S}/(w_{S}+w_{L})$ describing the mixing of the contributing antenna's dimensions. The black circles and stars in Fig.~\ref{fig:4} are the data point obtained from 1950 nm and 830 nm arm length, respectively [Fig.~\ref{fig:2} and Fig.~\ref{fig:3}]. The quantities $w_{S}$ and $w_{L}$ are inferred by comparing and matching the respective experimental and theoretical Fourier planes. It becomes apparent that the black and orange markers form a single trend. This correlation is a demonstration of the SHG Fourier plane imaging potency to monitor near-field coupling in plasmonic systems, and, in conjunction with theoretical modelling, to constitute a tool for nonlinear optical far-field reconstruction of plasmonic near-fields.

In summary, by implementing spectrally filtered Fourier and image plane measurements and a simple phenomenological model, we demonstrate that nonlinear responses in gold rod optical antennas can have a significant spatially delocalized contribution when excited locally by a focused laser beam. This result presents an alternative - delocalized - mechanism of nonlinear emission excitation in plasmonic structures, which can be adopted to observe and explain a variety of parametric and non-parametric nonlinear processes. We anticipate that incorporation of the delocalization concept in the current nonlinear plasmonic discourse can incite novel approaches for coherent control and tailoring of nonlinearities in single plasmonic cavities and metamaterials via plasmon mediated (quasi) phase matching.

The research leading to these results has received fundings from the European Research Council under the European Community's 7th Framework Program FP7/2007–2013 Grant Agreement no 306772 and the Agence Nationale de la Recherche (grants PLACORE ANR-13-BS10-0007). O. D. and B. C. thank the Labex ACTION program (contract ANR-11-LABX-0001-01). The authors thank G. Bachelier for stimulating discussions.


\begin{thebibliography}{50}%
\makeatletter
\providecommand \@ifxundefined [1]{%
 \@ifx{#1\undefined}
}%
\providecommand \@ifnum [1]{%
 \ifnum #1\expandafter \@firstoftwo
 \else \expandafter \@secondoftwo
 \fi
}%
\providecommand \@ifx [1]{%
 \ifx #1\expandafter \@firstoftwo
 \else \expandafter \@secondoftwo
 \fi
}%
\providecommand \natexlab [1]{#1}%
\providecommand \enquote  [1]{``#1''}%
\providecommand \bibnamefont  [1]{#1}%
\providecommand \bibfnamefont [1]{#1}%
\providecommand \citenamefont [1]{#1}%
\providecommand \href@noop [0]{\@secondoftwo}%
\providecommand \href [0]{\begingroup \@sanitize@url \@href}%
\providecommand \@href[1]{\@@startlink{#1}\@@href}%
\providecommand \@@href[1]{\endgroup#1\@@endlink}%
\providecommand \@sanitize@url [0]{\catcode `\\12\catcode `\$12\catcode
  `\&amp;12\catcode `\#12\catcode `\^12\catcode `\_12\catcode `\%12\relax}%
\providecommand \@@startlink[1]{}%
\providecommand \@@endlink[0]{}%
\providecommand \url  [0]{\begingroup\@sanitize@url \@url }%
\providecommand \@url [1]{\endgroup\@href {#1}{\urlprefix }}%
\providecommand \urlprefix  [0]{URL }%
\providecommand \Eprint [0]{\href }%
\providecommand \doibase [0]{http://dx.doi.org/}%
\providecommand \selectlanguage [0]{\@gobble}%
\providecommand \bibinfo  [0]{\@secondoftwo}%
\providecommand \bibfield  [0]{\@secondoftwo}%
\providecommand \translation [1]{[#1]}%
\providecommand \BibitemOpen [0]{}%
\providecommand \bibitemStop [0]{}%
\providecommand \bibitemNoStop [0]{.\EOS\space}%
\providecommand \EOS [0]{\spacefactor3000\relax}%
\providecommand \BibitemShut  [1]{\csname bibitem#1\endcsname}%
\let\auto@bib@innerbib\@empty
\bibitem [{\citenamefont {Schuller}\ \emph {et~al.}(2010)\citenamefont
  {Schuller}, \citenamefont {Barnard}, \citenamefont {Cai}, \citenamefont
  {Jun}, \citenamefont {White},\ and\ \citenamefont {Brongersma}}]{Schuller10}%
  \BibitemOpen
  \bibfield  {author} {\bibinfo {author} {\bibfnamefont {J.~A.}\ \bibnamefont
  {Schuller}}, \bibinfo {author} {\bibfnamefont {E.~S.}\ \bibnamefont
  {Barnard}}, \bibinfo {author} {\bibfnamefont {W.}~\bibnamefont {Cai}},
  \bibinfo {author} {\bibfnamefont {Y.~C.}\ \bibnamefont {Jun}}, \bibinfo
  {author} {\bibfnamefont {J.~S.}\ \bibnamefont {White}}, \ and\ \bibinfo
  {author} {\bibfnamefont {M.~L.}\ \bibnamefont {Brongersma}},\ }\href
  {http://dx.doi.org/10.1038/nmat2630} {\bibfield  {journal} {\bibinfo
  {journal} {Nature Mat.}\ }\textbf {\bibinfo {volume} {9}},\ \bibinfo {pages}
  {193} (\bibinfo {year} {2010})}\BibitemShut {NoStop}%
\bibitem [{\citenamefont {Novotny}\ and\ \citenamefont
  {Van~Hulst}(2011)}]{novotny11NP}%
  \BibitemOpen
  \bibfield  {author} {\bibinfo {author} {\bibfnamefont {L.}~\bibnamefont
  {Novotny}}\ and\ \bibinfo {author} {\bibfnamefont {N.~F.}\ \bibnamefont
  {Van~Hulst}},\ }\href@noop {} {\bibfield  {journal} {\bibinfo  {journal}
  {Nature Phot.}\ }\textbf {\bibinfo {volume} {5}},\ \bibinfo {pages} {83}
  (\bibinfo {year} {2011})}\BibitemShut {NoStop}%
\bibitem [{\citenamefont {Agio}\ and\ \citenamefont
  {Al{\`u}}(2013)}]{OptAntBook13}%
  \BibitemOpen
  \bibinfo {editor} {\bibfnamefont {M.}~\bibnamefont {Agio}}\ and\ \bibinfo
  {editor} {\bibfnamefont {A.}~\bibnamefont {Al{\`u}}},\ eds.,\ \href@noop {}
  {\emph {\bibinfo {title} {Optical Antennas}}}\ (\bibinfo  {publisher}
  {Cambridge University Press},\ \bibinfo {address} {Cambridge, United
  Kingdom},\ \bibinfo {year} {2013})\BibitemShut {NoStop}%
\bibitem [{\citenamefont {Kauranen}\ and\ \citenamefont
  {Zayats}(2012)}]{Kauranen:2012aa}%
  \BibitemOpen
  \bibfield  {author} {\bibinfo {author} {\bibfnamefont {M.}~\bibnamefont
  {Kauranen}}\ and\ \bibinfo {author} {\bibfnamefont {A.~V.}\ \bibnamefont
  {Zayats}},\ }\href {http://dx.doi.org/10.1038/nphoton.2012.244} {\bibfield
  {journal} {\bibinfo  {journal} {Nature Phot.}\ }\textbf {\bibinfo {volume}
  {6}},\ \bibinfo {pages} {737} (\bibinfo {year} {2012})}\BibitemShut {NoStop}%
\bibitem [{\citenamefont {Harutyunyan}\ \emph {et~al.}(2013)\citenamefont
  {Harutyunyan}, \citenamefont {Volpe},\ and\ \citenamefont
  {Novonty}}]{NovotnyNLbook}%
  \BibitemOpen
  \bibfield  {author} {\bibinfo {author} {\bibfnamefont {H.}~\bibnamefont
  {Harutyunyan}}, \bibinfo {author} {\bibfnamefont {G.}~\bibnamefont {Volpe}},
  \ and\ \bibinfo {author} {\bibfnamefont {L.}~\bibnamefont {Novonty}},\
  }\enquote {\bibinfo {title} {Optical antennas},}\ \ (\bibinfo  {publisher}
  {Cambridge University Press},\ \bibinfo {year} {2013})\ Chap.\ \bibinfo
  {chapter} {Nonlinear optical antennas}, pp.\ \bibinfo {pages}
  {131--144}\BibitemShut {NoStop}%
\bibitem [{\citenamefont {Ginzburg}\ and\ \citenamefont
  {Orenstein}(2013)}]{NLObook}%
  \BibitemOpen
  \bibfield  {author} {\bibinfo {author} {\bibfnamefont {P.}~\bibnamefont
  {Ginzburg}}\ and\ \bibinfo {author} {\bibfnamefont {M.}~\bibnamefont
  {Orenstein}},\ }\enquote {\bibinfo {title} {Nonlinear effects in plasmonic
  systems},}\ in\ \href {\doibase 10.1002/9781118634394.ch2} {\emph {\bibinfo
  {booktitle} {Active Plasmonics and Tuneable Plasmonic Metamaterials}}}\
  (\bibinfo  {publisher} {John Wiley \& Sons, Inc.},\ \bibinfo {year} {2013})\
  pp.\ \bibinfo {pages} {41--67}\BibitemShut {NoStop}%
\bibitem [{\citenamefont {Kim}\ \emph {et~al.}(2008)\citenamefont {Kim},
  \citenamefont {Jin}, \citenamefont {Kim}, \citenamefont {Park}, \citenamefont
  {Kim},\ and\ \citenamefont {Kim}}]{kimNature08}%
  \BibitemOpen
  \bibfield  {author} {\bibinfo {author} {\bibfnamefont {S.}~\bibnamefont
  {Kim}}, \bibinfo {author} {\bibfnamefont {J.}~\bibnamefont {Jin}}, \bibinfo
  {author} {\bibfnamefont {Y.-J.}\ \bibnamefont {Kim}}, \bibinfo {author}
  {\bibfnamefont {I.-Y.}\ \bibnamefont {Park}}, \bibinfo {author}
  {\bibfnamefont {Y.}~\bibnamefont {Kim}}, \ and\ \bibinfo {author}
  {\bibfnamefont {S.-W.}\ \bibnamefont {Kim}},\ }\href@noop {} {\bibfield
  {journal} {\bibinfo  {journal} {Nature}\ }\textbf {\bibinfo {volume} {453}},\
  \bibinfo {pages} {757} (\bibinfo {year} {2008})}\BibitemShut {NoStop}%
\bibitem [{\citenamefont {Stolz}\ \emph {et~al.}(2014)\citenamefont {Stolz},
  \citenamefont {Berthelot}, \citenamefont {Mennemanteuil}, \citenamefont
  {{Colas des Francs}}, \citenamefont {Markey}, \citenamefont {Meunier},\ and\
  \citenamefont {Bouhelier}}]{Stolz2014}%
  \BibitemOpen
  \bibfield  {author} {\bibinfo {author} {\bibfnamefont {A.}~\bibnamefont
  {Stolz}}, \bibinfo {author} {\bibfnamefont {J.}~\bibnamefont {Berthelot}},
  \bibinfo {author} {\bibfnamefont {M.-M.}\ \bibnamefont {Mennemanteuil}},
  \bibinfo {author} {\bibfnamefont {G.}~\bibnamefont {{Colas des Francs}}},
  \bibinfo {author} {\bibfnamefont {L.}~\bibnamefont {Markey}}, \bibinfo
  {author} {\bibfnamefont {V.}~\bibnamefont {Meunier}}, \ and\ \bibinfo
  {author} {\bibfnamefont {A.}~\bibnamefont {Bouhelier}},\ }\href@noop {}
  {\bibfield  {journal} {\bibinfo  {journal} {Nano Lett.}\ }\textbf {\bibinfo
  {volume} {14}},\ \bibinfo {pages} {1} (\bibinfo {year} {2014})}\BibitemShut
  {NoStop}%
\bibitem [{\citenamefont {Abb}\ \emph {et~al.}(2011)\citenamefont {Abb},
  \citenamefont {Albella}, \citenamefont {Aizpurua},\ and\ \citenamefont
  {Muskens}}]{Muskens11}%
  \BibitemOpen
  \bibfield  {author} {\bibinfo {author} {\bibfnamefont {M.}~\bibnamefont
  {Abb}}, \bibinfo {author} {\bibfnamefont {P.}~\bibnamefont {Albella}},
  \bibinfo {author} {\bibfnamefont {J.}~\bibnamefont {Aizpurua}}, \ and\
  \bibinfo {author} {\bibfnamefont {O.~L.}\ \bibnamefont {Muskens}},\ }\href
  {\doibase 10.1021/nl200901w} {\bibfield  {journal} {\bibinfo  {journal} {Nano
  Lett.}\ }\textbf {\bibinfo {volume} {11}},\ \bibinfo {pages} {2457} (\bibinfo
  {year} {2011})}\BibitemShut {NoStop}%
\bibitem [{\citenamefont {Wurtz}\ \emph {et~al.}(2011)\citenamefont {Wurtz},
  \citenamefont {Pollard}, \citenamefont {Hendren}, \citenamefont
  {Wiederrecht}, \citenamefont {Gosztola}, \citenamefont {Podolskiy},\ and\
  \citenamefont {Zayats}}]{Wurtz11}%
  \BibitemOpen
  \bibfield  {author} {\bibinfo {author} {\bibfnamefont {G.~A.}\ \bibnamefont
  {Wurtz}}, \bibinfo {author} {\bibfnamefont {R.}~\bibnamefont {Pollard}},
  \bibinfo {author} {\bibfnamefont {W.}~\bibnamefont {Hendren}}, \bibinfo
  {author} {\bibfnamefont {G.~P.}\ \bibnamefont {Wiederrecht}}, \bibinfo
  {author} {\bibfnamefont {D.~J.}\ \bibnamefont {Gosztola}}, \bibinfo {author}
  {\bibfnamefont {V.~A.}\ \bibnamefont {Podolskiy}}, \ and\ \bibinfo {author}
  {\bibfnamefont {A.~V.}\ \bibnamefont {Zayats}},\ }\href@noop {} {\bibfield
  {journal} {\bibinfo  {journal} {Nature Nanotech.}\ }\textbf {\bibinfo
  {volume} {6}},\ \bibinfo {pages} {107} (\bibinfo {year} {2011})}\BibitemShut
  {NoStop}%
\bibitem [{\citenamefont {De~Leon}\ \emph {et~al.}(2014)\citenamefont
  {De~Leon}, \citenamefont {Sipe},\ and\ \citenamefont {Boyd}}]{DeLeon:14}%
  \BibitemOpen
  \bibfield  {author} {\bibinfo {author} {\bibfnamefont {I.}~\bibnamefont
  {De~Leon}}, \bibinfo {author} {\bibfnamefont {J.~E.}\ \bibnamefont {Sipe}}, \
  and\ \bibinfo {author} {\bibfnamefont {R.~W.}\ \bibnamefont {Boyd}},\ }\href
  {\doibase 10.1103/PhysRevA.89.013855} {\bibfield  {journal} {\bibinfo
  {journal} {Phys. Rev. A}\ }\textbf {\bibinfo {volume} {89}},\ \bibinfo
  {pages} {013855} (\bibinfo {year} {2014})}\BibitemShut {NoStop}%
\bibitem [{\citenamefont {Baron}\ \emph {et~al.}(2015)\citenamefont {Baron},
  \citenamefont {Larouche}, \citenamefont {Gauthier},\ and\ \citenamefont
  {Smith}}]{Baron:15}%
  \BibitemOpen
  \bibfield  {author} {\bibinfo {author} {\bibfnamefont {A.}~\bibnamefont
  {Baron}}, \bibinfo {author} {\bibfnamefont {S.}~\bibnamefont {Larouche}},
  \bibinfo {author} {\bibfnamefont {D.~J.}\ \bibnamefont {Gauthier}}, \ and\
  \bibinfo {author} {\bibfnamefont {D.~R.}\ \bibnamefont {Smith}},\ }\href
  {\doibase 10.1364/JOSAB.32.000009} {\bibfield  {journal} {\bibinfo  {journal}
  {J. Opt. Soc. Am. B}\ }\textbf {\bibinfo {volume} {32}},\ \bibinfo {pages}
  {9} (\bibinfo {year} {2015})}\BibitemShut {NoStop}%
\bibitem [{\citenamefont {Dadap}\ \emph {et~al.}(2004)\citenamefont {Dadap},
  \citenamefont {Shan},\ and\ \citenamefont {Heinz}}]{Dadap04}%
  \BibitemOpen
  \bibfield  {author} {\bibinfo {author} {\bibfnamefont {J.~I.}\ \bibnamefont
  {Dadap}}, \bibinfo {author} {\bibfnamefont {J.}~\bibnamefont {Shan}}, \ and\
  \bibinfo {author} {\bibfnamefont {T.~F.}\ \bibnamefont {Heinz}},\ }\href
  {\doibase 10.1364/JOSAB.21.001328} {\bibfield  {journal} {\bibinfo  {journal}
  {J. Opt. Soc. Am. B}\ }\textbf {\bibinfo {volume} {21}},\ \bibinfo {pages}
  {1328} (\bibinfo {year} {2004})}\BibitemShut {NoStop}%
\bibitem [{\citenamefont {Lippitz}\ \emph {et~al.}(2005)\citenamefont
  {Lippitz}, \citenamefont {van Dijk},\ and\ \citenamefont
  {Orrit}}]{orrit05NL}%
  \BibitemOpen
  \bibfield  {author} {\bibinfo {author} {\bibfnamefont {M.}~\bibnamefont
  {Lippitz}}, \bibinfo {author} {\bibfnamefont {M.~A.}\ \bibnamefont {van
  Dijk}}, \ and\ \bibinfo {author} {\bibfnamefont {M.}~\bibnamefont {Orrit}},\
  }\href@noop {} {\bibfield  {journal} {\bibinfo  {journal} {Nano Lett.}\
  }\textbf {\bibinfo {volume} {5}},\ \bibinfo {pages} {799} (\bibinfo {year}
  {2005})}\BibitemShut {NoStop}%
\bibitem [{\citenamefont {Bouhelier}\ \emph {et~al.}(2005)\citenamefont
  {Bouhelier}, \citenamefont {Bachelot}, \citenamefont {Lerondel},
  \citenamefont {Kostcheev}, \citenamefont {Royer},\ and\ \citenamefont
  {Wiederrecht}}]{bouhelier05PRL}%
  \BibitemOpen
  \bibfield  {author} {\bibinfo {author} {\bibfnamefont {A.}~\bibnamefont
  {Bouhelier}}, \bibinfo {author} {\bibfnamefont {R.}~\bibnamefont {Bachelot}},
  \bibinfo {author} {\bibfnamefont {G.}~\bibnamefont {Lerondel}}, \bibinfo
  {author} {\bibfnamefont {S.}~\bibnamefont {Kostcheev}}, \bibinfo {author}
  {\bibfnamefont {P.}~\bibnamefont {Royer}}, \ and\ \bibinfo {author}
  {\bibfnamefont {G.~P.}\ \bibnamefont {Wiederrecht}},\ }\href@noop {}
  {\bibfield  {journal} {\bibinfo  {journal} {Phys. Rev. Lett.}\ }\textbf
  {\bibinfo {volume} {95}},\ \bibinfo {pages} {267405} (\bibinfo {year}
  {2005})}\BibitemShut {NoStop}%
\bibitem [{\citenamefont {Bachelier}\ \emph {et~al.}(2008)\citenamefont
  {Bachelier}, \citenamefont {Russier-Antoine}, \citenamefont {Benichou},
  \citenamefont {Jonin},\ and\ \citenamefont {Brevet}}]{Bachelier08}%
  \BibitemOpen
  \bibfield  {author} {\bibinfo {author} {\bibfnamefont {G.}~\bibnamefont
  {Bachelier}}, \bibinfo {author} {\bibfnamefont {I.}~\bibnamefont
  {Russier-Antoine}}, \bibinfo {author} {\bibfnamefont {E.}~\bibnamefont
  {Benichou}}, \bibinfo {author} {\bibfnamefont {C.}~\bibnamefont {Jonin}}, \
  and\ \bibinfo {author} {\bibfnamefont {P.-F.}\ \bibnamefont {Brevet}},\
  }\href@noop {} {\bibfield  {journal} {\bibinfo  {journal} {J. Opt. Soc. Am.
  B}\ }\textbf {\bibinfo {volume} {25}},\ \bibinfo {pages} {955} (\bibinfo
  {year} {2008})}\BibitemShut {NoStop}%
\bibitem [{\citenamefont {Wang}\ \emph {et~al.}(2009)\citenamefont {Wang},
  \citenamefont {Rodr\'iguez}, \citenamefont {Albers}, \citenamefont
  {Ahorinta}, \citenamefont {Sipe},\ and\ \citenamefont {Kauranen}}]{sipe09}%
  \BibitemOpen
  \bibfield  {author} {\bibinfo {author} {\bibfnamefont {F.~X.}\ \bibnamefont
  {Wang}}, \bibinfo {author} {\bibfnamefont {F.~J.}\ \bibnamefont
  {Rodr\'iguez}}, \bibinfo {author} {\bibfnamefont {W.~M.}\ \bibnamefont
  {Albers}}, \bibinfo {author} {\bibfnamefont {R.}~\bibnamefont {Ahorinta}},
  \bibinfo {author} {\bibfnamefont {J.~E.}\ \bibnamefont {Sipe}}, \ and\
  \bibinfo {author} {\bibfnamefont {M.}~\bibnamefont {Kauranen}},\ }\href
  {\doibase 10.1103/PhysRevB.80.233402} {\bibfield  {journal} {\bibinfo
  {journal} {Phys. Rev. B}\ }\textbf {\bibinfo {volume} {80}},\ \bibinfo
  {pages} {233402} (\bibinfo {year} {2009})}\BibitemShut {NoStop}%
\bibitem [{\citenamefont {Danckwerts}\ and\ \citenamefont
  {Novotny}(2007)}]{danckwerts07}%
  \BibitemOpen
  \bibfield  {author} {\bibinfo {author} {\bibfnamefont {M.}~\bibnamefont
  {Danckwerts}}\ and\ \bibinfo {author} {\bibfnamefont {L.}~\bibnamefont
  {Novotny}},\ }\href@noop {} {\bibfield  {journal} {\bibinfo  {journal} {Phys.
  Rev. Lett.}\ }\textbf {\bibinfo {volume} {98}},\ \bibinfo {pages} {026104}
  (\bibinfo {year} {2007})}\BibitemShut {NoStop}%
\bibitem [{\citenamefont {Ginzburg}\ \emph {et~al.}(2010)\citenamefont
  {Ginzburg}, \citenamefont {Hayat}, \citenamefont {Berkovitch},\ and\
  \citenamefont {Orenstein}}]{Ginzburg10}%
  \BibitemOpen
  \bibfield  {author} {\bibinfo {author} {\bibfnamefont {P.}~\bibnamefont
  {Ginzburg}}, \bibinfo {author} {\bibfnamefont {A.}~\bibnamefont {Hayat}},
  \bibinfo {author} {\bibfnamefont {N.}~\bibnamefont {Berkovitch}}, \ and\
  \bibinfo {author} {\bibfnamefont {M.}~\bibnamefont {Orenstein}},\ }\href
  {\doibase 10.1364/OL.35.001551} {\bibfield  {journal} {\bibinfo  {journal}
  {Opt. Lett.}\ }\textbf {\bibinfo {volume} {35}},\ \bibinfo {pages} {1551}
  (\bibinfo {year} {2010})}\BibitemShut {NoStop}%
\bibitem [{\citenamefont {Cai}\ \emph {et~al.}(2011)\citenamefont {Cai},
  \citenamefont {Vasudev},\ and\ \citenamefont {Brongersma}}]{Cai11}%
  \BibitemOpen
  \bibfield  {author} {\bibinfo {author} {\bibfnamefont {W.}~\bibnamefont
  {Cai}}, \bibinfo {author} {\bibfnamefont {A.~P.}\ \bibnamefont {Vasudev}}, \
  and\ \bibinfo {author} {\bibfnamefont {M.~L.}\ \bibnamefont {Brongersma}},\
  }\href {\doibase 10.1126/science.1207858} {\bibfield  {journal} {\bibinfo
  {journal} {Science}\ }\textbf {\bibinfo {volume} {333}},\ \bibinfo {pages}
  {1720} (\bibinfo {year} {2011})}\BibitemShut {NoStop}%
\bibitem [{\citenamefont {Metzger}\ \emph {et~al.}(2014)\citenamefont
  {Metzger}, \citenamefont {Hentschel}, \citenamefont {Schumacher},
  \citenamefont {Lippitz}, \citenamefont {Ye}, \citenamefont {Murray},
  \citenamefont {Knabe}, \citenamefont {Buse},\ and\ \citenamefont
  {Giessen}}]{metzger14}%
  \BibitemOpen
  \bibfield  {author} {\bibinfo {author} {\bibfnamefont {B.}~\bibnamefont
  {Metzger}}, \bibinfo {author} {\bibfnamefont {M.}~\bibnamefont {Hentschel}},
  \bibinfo {author} {\bibfnamefont {T.}~\bibnamefont {Schumacher}}, \bibinfo
  {author} {\bibfnamefont {M.}~\bibnamefont {Lippitz}}, \bibinfo {author}
  {\bibfnamefont {X.}~\bibnamefont {Ye}}, \bibinfo {author} {\bibfnamefont
  {C.~B.}\ \bibnamefont {Murray}}, \bibinfo {author} {\bibfnamefont
  {B.}~\bibnamefont {Knabe}}, \bibinfo {author} {\bibfnamefont
  {K.}~\bibnamefont {Buse}}, \ and\ \bibinfo {author} {\bibfnamefont
  {H.}~\bibnamefont {Giessen}},\ }\href {\doibase 10.1021/nl500913t} {\bibfield
   {journal} {\bibinfo  {journal} {Nano Lett.}\ }\textbf {\bibinfo {volume}
  {14}},\ \bibinfo {pages} {2867} (\bibinfo {year} {2014})}\BibitemShut
  {NoStop}%
\bibitem [{\citenamefont {Ginzburg}\ \emph
  {et~al.}(2015{\natexlab{a}})\citenamefont {Ginzburg}, \citenamefont
  {Krasavin}, \citenamefont {Wurtz},\ and\ \citenamefont
  {Zayats}}]{ginsburg14}%
  \BibitemOpen
  \bibfield  {author} {\bibinfo {author} {\bibfnamefont {P.}~\bibnamefont
  {Ginzburg}}, \bibinfo {author} {\bibfnamefont {A.~V.}\ \bibnamefont
  {Krasavin}}, \bibinfo {author} {\bibfnamefont {G.~A.}\ \bibnamefont {Wurtz}},
  \ and\ \bibinfo {author} {\bibfnamefont {A.~V.}\ \bibnamefont {Zayats}},\
  }\href {\doibase 10.1021/ph500362y} {\bibfield  {journal} {\bibinfo
  {journal} {ACS Photonics}\ }\textbf {\bibinfo {volume} {2}},\ \bibinfo
  {pages} {8} (\bibinfo {year} {2015}{\natexlab{a}})}\BibitemShut {NoStop}%
\bibitem [{\citenamefont {M\"uhlschlegel}\ \emph {et~al.}(2005)\citenamefont
  {M\"uhlschlegel}, \citenamefont {Eisler}, \citenamefont {Martin},
  \citenamefont {Hecht},\ and\ \citenamefont {Pohl}}]{hecht05sciences}%
  \BibitemOpen
  \bibfield  {author} {\bibinfo {author} {\bibfnamefont {P.}~\bibnamefont
  {M\"uhlschlegel}}, \bibinfo {author} {\bibfnamefont {H.-J.}\ \bibnamefont
  {Eisler}}, \bibinfo {author} {\bibfnamefont {O.~J.~F.}\ \bibnamefont
  {Martin}}, \bibinfo {author} {\bibfnamefont {B.}~\bibnamefont {Hecht}}, \
  and\ \bibinfo {author} {\bibfnamefont {D.~W.}\ \bibnamefont {Pohl}},\
  }\href@noop {} {\bibfield  {journal} {\bibinfo  {journal} {Science}\ }\textbf
  {\bibinfo {volume} {308}},\ \bibinfo {pages} {1607} (\bibinfo {year}
  {2005})}\BibitemShut {NoStop}%
\bibitem [{\citenamefont {Harutyunyan}\ \emph {et~al.}(2012)\citenamefont
  {Harutyunyan}, \citenamefont {Volpe}, \citenamefont {Quidant},\ and\
  \citenamefont {Novotny}}]{Novotny_11}%
  \BibitemOpen
  \bibfield  {author} {\bibinfo {author} {\bibfnamefont {H.}~\bibnamefont
  {Harutyunyan}}, \bibinfo {author} {\bibfnamefont {G.}~\bibnamefont {Volpe}},
  \bibinfo {author} {\bibfnamefont {R.}~\bibnamefont {Quidant}}, \ and\
  \bibinfo {author} {\bibfnamefont {L.}~\bibnamefont {Novotny}},\ }\href@noop
  {} {\bibfield  {journal} {\bibinfo  {journal} {Phys. Rev. Lett.}\ }\textbf
  {\bibinfo {volume} {108}},\ \bibinfo {pages} {217403} (\bibinfo {year}
  {2012})}\BibitemShut {NoStop}%
\bibitem [{\citenamefont {Knittel}\ \emph {et~al.}(2015)\citenamefont
  {Knittel}, \citenamefont {Fischer}, \citenamefont {de~Roo}, \citenamefont
  {Mecking}, \citenamefont {Leitenstorfer},\ and\ \citenamefont
  {Brida}}]{Knittel:15}%
  \BibitemOpen
  \bibfield  {author} {\bibinfo {author} {\bibfnamefont {V.}~\bibnamefont
  {Knittel}}, \bibinfo {author} {\bibfnamefont {M.~P.}\ \bibnamefont
  {Fischer}}, \bibinfo {author} {\bibfnamefont {T.}~\bibnamefont {de~Roo}},
  \bibinfo {author} {\bibfnamefont {S.}~\bibnamefont {Mecking}}, \bibinfo
  {author} {\bibfnamefont {A.}~\bibnamefont {Leitenstorfer}}, \ and\ \bibinfo
  {author} {\bibfnamefont {D.}~\bibnamefont {Brida}},\ }\href {\doibase
  10.1021/nn5066233} {\bibfield  {journal} {\bibinfo  {journal} {ACS Nano}\
  }\textbf {\bibinfo {volume} {9}},\ \bibinfo {pages} {894} (\bibinfo {year}
  {2015})}\BibitemShut {NoStop}%
\bibitem [{\citenamefont {Ghenuche}\ \emph {et~al.}(2008)\citenamefont
  {Ghenuche}, \citenamefont {Cherukulappurath}, \citenamefont {Taminiau},
  \citenamefont {F.{ van Hulst}},\ and\ \citenamefont {Quidant}}]{quidant08}%
  \BibitemOpen
  \bibfield  {author} {\bibinfo {author} {\bibfnamefont {P.}~\bibnamefont
  {Ghenuche}}, \bibinfo {author} {\bibfnamefont {S.}~\bibnamefont
  {Cherukulappurath}}, \bibinfo {author} {\bibfnamefont {T.~H.}\ \bibnamefont
  {Taminiau}}, \bibinfo {author} {\bibfnamefont {N.}~\bibnamefont {F.{ van
  Hulst}}}, \ and\ \bibinfo {author} {\bibfnamefont {R.}~\bibnamefont
  {Quidant}},\ }\href@noop {} {\bibfield  {journal} {\bibinfo  {journal} {Phys.
  Rev. Lett.}\ }\textbf {\bibinfo {volume} {101}},\ \bibinfo {pages} {116805}
  (\bibinfo {year} {2008})}\BibitemShut {NoStop}%
\bibitem [{\citenamefont {Berthelot}\ \emph {et~al.}(2012)\citenamefont
  {Berthelot}, \citenamefont {Tantussi}, \citenamefont {Rai}, \citenamefont
  {{Colas des Francs}}, \citenamefont {Weeber}, \citenamefont {Dereux},
  \citenamefont {Fuso}, \citenamefont {Allegrini},\ and\ \citenamefont
  {Bouhelier}}]{Berthelot:12}%
  \BibitemOpen
  \bibfield  {author} {\bibinfo {author} {\bibfnamefont {J.}~\bibnamefont
  {Berthelot}}, \bibinfo {author} {\bibfnamefont {F.}~\bibnamefont {Tantussi}},
  \bibinfo {author} {\bibfnamefont {P.}~\bibnamefont {Rai}}, \bibinfo {author}
  {\bibfnamefont {G.}~\bibnamefont {{Colas des Francs}}}, \bibinfo {author}
  {\bibfnamefont {J.-C.}\ \bibnamefont {Weeber}}, \bibinfo {author}
  {\bibfnamefont {A.}~\bibnamefont {Dereux}}, \bibinfo {author} {\bibfnamefont
  {F.}~\bibnamefont {Fuso}}, \bibinfo {author} {\bibfnamefont {M.}~\bibnamefont
  {Allegrini}}, \ and\ \bibinfo {author} {\bibfnamefont {A.}~\bibnamefont
  {Bouhelier}},\ }\href {\doibase 10.1364/JOSAB.29.000226} {\bibfield
  {journal} {\bibinfo  {journal} {J. Opt. Soc. Am. B}\ }\textbf {\bibinfo
  {volume} {29}},\ \bibinfo {pages} {226} (\bibinfo {year} {2012})}\BibitemShut
  {NoStop}%
\bibitem [{\citenamefont {Valev}\ \emph {et~al.}(2012)\citenamefont {Valev},
  \citenamefont {Clercq}, \citenamefont {Zheng}, \citenamefont {Denkova},
  \citenamefont {Osley}, \citenamefont {Vandendriessche}, \citenamefont
  {Silhanek}, \citenamefont {Volskiy}, \citenamefont {Warburton}, \citenamefont
  {Vandenbosch},\ and\ \citenamefont {et~al.}}]{valev:12}%
  \BibitemOpen
  \bibfield  {author} {\bibinfo {author} {\bibfnamefont {V.~K.}\ \bibnamefont
  {Valev}}, \bibinfo {author} {\bibfnamefont {B.~D.}\ \bibnamefont {Clercq}},
  \bibinfo {author} {\bibfnamefont {X.}~\bibnamefont {Zheng}}, \bibinfo
  {author} {\bibfnamefont {D.}~\bibnamefont {Denkova}}, \bibinfo {author}
  {\bibfnamefont {E.~J.}\ \bibnamefont {Osley}}, \bibinfo {author}
  {\bibfnamefont {S.}~\bibnamefont {Vandendriessche}}, \bibinfo {author}
  {\bibfnamefont {A.~V.}\ \bibnamefont {Silhanek}}, \bibinfo {author}
  {\bibfnamefont {V.}~\bibnamefont {Volskiy}}, \bibinfo {author} {\bibfnamefont
  {P.~A.}\ \bibnamefont {Warburton}}, \bibinfo {author} {\bibfnamefont
  {G.~A.~E.}\ \bibnamefont {Vandenbosch}}, \ and\ \bibinfo {author}
  {\bibnamefont {et~al.}},\ }\href {\doibase 10.1364/OE.20.000256} {\bibfield
  {journal} {\bibinfo  {journal} {Opt. Express}\ }\textbf {\bibinfo {volume}
  {20}},\ \bibinfo {pages} {256} (\bibinfo {year} {2012})}\BibitemShut
  {NoStop}%
\bibitem [{\citenamefont {Viarbitskaya}\ \emph {et~al.}(2013)\citenamefont
  {Viarbitskaya}, \citenamefont {Teulle}, \citenamefont {Marty}, \citenamefont
  {Sharma}, \citenamefont {Girard}, \citenamefont {Arbouet},\ and\
  \citenamefont {Dujardin}}]{viarb13nm}%
  \BibitemOpen
  \bibfield  {author} {\bibinfo {author} {\bibfnamefont {S.}~\bibnamefont
  {Viarbitskaya}}, \bibinfo {author} {\bibfnamefont {A.}~\bibnamefont
  {Teulle}}, \bibinfo {author} {\bibfnamefont {R.}~\bibnamefont {Marty}},
  \bibinfo {author} {\bibfnamefont {J.}~\bibnamefont {Sharma}}, \bibinfo
  {author} {\bibfnamefont {C.}~\bibnamefont {Girard}}, \bibinfo {author}
  {\bibfnamefont {A.}~\bibnamefont {Arbouet}}, \ and\ \bibinfo {author}
  {\bibfnamefont {E.}~\bibnamefont {Dujardin}},\ }\href
  {http://dx.doi.org/10.1038/nmat3581} {\bibfield  {journal} {\bibinfo
  {journal} {Nature Mat.}\ }\textbf {\bibinfo {volume} {12}},\ \bibinfo {pages}
  {426} (\bibinfo {year} {2013})}\BibitemShut {NoStop}%
\bibitem [{\citenamefont {Demichel}\ \emph {et~al.}(2014)\citenamefont
  {Demichel}, \citenamefont {Petit}, \citenamefont {{Colas des Francs}},
  \citenamefont {Bouhelier}, \citenamefont {Hertz}, \citenamefont {Billard},
  \citenamefont {de~Fornel},\ and\ \citenamefont {Cluzel}}]{Demichel:14}%
  \BibitemOpen
  \bibfield  {author} {\bibinfo {author} {\bibfnamefont {O.}~\bibnamefont
  {Demichel}}, \bibinfo {author} {\bibfnamefont {M.}~\bibnamefont {Petit}},
  \bibinfo {author} {\bibfnamefont {G.}~\bibnamefont {{Colas des Francs}}},
  \bibinfo {author} {\bibfnamefont {A.}~\bibnamefont {Bouhelier}}, \bibinfo
  {author} {\bibfnamefont {E.}~\bibnamefont {Hertz}}, \bibinfo {author}
  {\bibfnamefont {F.}~\bibnamefont {Billard}}, \bibinfo {author} {\bibfnamefont
  {F.}~\bibnamefont {de~Fornel}}, \ and\ \bibinfo {author} {\bibfnamefont
  {B.}~\bibnamefont {Cluzel}},\ }\href {\doibase 10.1364/OE.22.015088}
  {\bibfield  {journal} {\bibinfo  {journal} {Opt. Express}\ }\textbf {\bibinfo
  {volume} {22}},\ \bibinfo {pages} {15088} (\bibinfo {year}
  {2014})}\BibitemShut {NoStop}%
\bibitem [{\citenamefont {Song}\ \emph {et~al.}(2013)\citenamefont {Song},
  \citenamefont {Stolz}, \citenamefont {Zhang}, \citenamefont {Arocas},
  \citenamefont {Markey}, \citenamefont {{Colas des Francs}}, \citenamefont
  {Dujardin},\ and\ \citenamefont {Bouhelier}}]{SongJOVE}%
  \BibitemOpen
  \bibfield  {author} {\bibinfo {author} {\bibfnamefont {M.}~\bibnamefont
  {Song}}, \bibinfo {author} {\bibfnamefont {A.}~\bibnamefont {Stolz}},
  \bibinfo {author} {\bibfnamefont {D.}~\bibnamefont {Zhang}}, \bibinfo
  {author} {\bibfnamefont {J.}~\bibnamefont {Arocas}}, \bibinfo {author}
  {\bibfnamefont {L.}~\bibnamefont {Markey}}, \bibinfo {author} {\bibfnamefont
  {G.}~\bibnamefont {{Colas des Francs}}}, \bibinfo {author} {\bibfnamefont
  {E.}~\bibnamefont {Dujardin}}, \ and\ \bibinfo {author} {\bibfnamefont
  {A.}~\bibnamefont {Bouhelier}},\ }\href@noop {} {\bibfield  {journal}
  {\bibinfo  {journal} {J. Vis. Exp.}\ }\textbf {\bibinfo {volume} {82}},\
  \bibinfo {pages} {e514048} (\bibinfo {year} {2013})}\BibitemShut {NoStop}%
\bibitem [{\citenamefont {Boyd}\ \emph {et~al.}(1986)\citenamefont {Boyd},
  \citenamefont {Yu},\ and\ \citenamefont {Shen}}]{boyd86}%
  \BibitemOpen
  \bibfield  {author} {\bibinfo {author} {\bibfnamefont {G.~T.}\ \bibnamefont
  {Boyd}}, \bibinfo {author} {\bibfnamefont {Z.~H.}\ \bibnamefont {Yu}}, \ and\
  \bibinfo {author} {\bibfnamefont {Y.~R.}\ \bibnamefont {Shen}},\ }\href@noop
  {} {\bibfield  {journal} {\bibinfo  {journal} {Phys. Rev. B}\ }\textbf
  {\bibinfo {volume} {33}},\ \bibinfo {pages} {7923} (\bibinfo {year}
  {1986})}\BibitemShut {NoStop}%
\bibitem [{\citenamefont {Biagioni}\ \emph {et~al.}(2012)\citenamefont
  {Biagioni}, \citenamefont {Brida}, \citenamefont {Huang}, \citenamefont
  {Kern}, \citenamefont {Duò}, \citenamefont {Hecht}, \citenamefont
  {Finazzi},\ and\ \citenamefont {Cerullo}}]{BiagoniNL12}%
  \BibitemOpen
  \bibfield  {author} {\bibinfo {author} {\bibfnamefont {P.}~\bibnamefont
  {Biagioni}}, \bibinfo {author} {\bibfnamefont {D.}~\bibnamefont {Brida}},
  \bibinfo {author} {\bibfnamefont {J.-S.}\ \bibnamefont {Huang}}, \bibinfo
  {author} {\bibfnamefont {J.}~\bibnamefont {Kern}}, \bibinfo {author}
  {\bibfnamefont {L.}~\bibnamefont {Duò}}, \bibinfo {author} {\bibfnamefont
  {B.}~\bibnamefont {Hecht}}, \bibinfo {author} {\bibfnamefont
  {M.}~\bibnamefont {Finazzi}}, \ and\ \bibinfo {author} {\bibfnamefont
  {G.}~\bibnamefont {Cerullo}},\ }\href@noop {} {\bibfield  {journal} {\bibinfo
   {journal} {Nano Lett.}\ }\textbf {\bibinfo {volume} {12}},\ \bibinfo {pages}
  {2941} (\bibinfo {year} {2012})}\BibitemShut {NoStop}%
\bibitem [{\citenamefont {Shahbazyan}(2013)}]{Tigran13}%
  \BibitemOpen
  \bibfield  {author} {\bibinfo {author} {\bibfnamefont {T.~V.}\ \bibnamefont
  {Shahbazyan}},\ }\href {\doibase 10.1021/nl303851z} {\bibfield  {journal}
  {\bibinfo  {journal} {Nano Lett.}\ }\textbf {\bibinfo {volume} {13}},\
  \bibinfo {pages} {194} (\bibinfo {year} {2013})}\BibitemShut {NoStop}%
\bibitem [{\citenamefont {Verellen}\ \emph {et~al.}(2015)\citenamefont
  {Verellen}, \citenamefont {Denkova}, \citenamefont {Clercq}, \citenamefont
  {Silhanek}, \citenamefont {Ameloot}, \citenamefont {Dorpe},\ and\
  \citenamefont {Moshchalkov}}]{verellen15}%
  \BibitemOpen
  \bibfield  {author} {\bibinfo {author} {\bibfnamefont {N.}~\bibnamefont
  {Verellen}}, \bibinfo {author} {\bibfnamefont {D.}~\bibnamefont {Denkova}},
  \bibinfo {author} {\bibfnamefont {B.~D.}\ \bibnamefont {Clercq}}, \bibinfo
  {author} {\bibfnamefont {A.~V.}\ \bibnamefont {Silhanek}}, \bibinfo {author}
  {\bibfnamefont {M.}~\bibnamefont {Ameloot}}, \bibinfo {author} {\bibfnamefont
  {P.~V.}\ \bibnamefont {Dorpe}}, \ and\ \bibinfo {author} {\bibfnamefont
  {V.~V.}\ \bibnamefont {Moshchalkov}},\ }\href {\doibase 10.1021/ph500453m}
  {\bibfield  {journal} {\bibinfo  {journal} {ACS Photonics}\ }\textbf
  {\bibinfo {volume} {2}},\ \bibinfo {pages} {410} (\bibinfo {year}
  {2015})}\BibitemShut {NoStop}%
\bibitem [{\citenamefont {Huang}\ \emph {et~al.}(2010)\citenamefont {Huang},
  \citenamefont {Kern}, \citenamefont {Geisler}, \citenamefont {Weinmann},
  \citenamefont {Kamp}, \citenamefont {Forchel}, \citenamefont {Biagioni},\
  and\ \citenamefont {Hecht.}}]{hechtNL10}%
  \BibitemOpen
  \bibfield  {author} {\bibinfo {author} {\bibfnamefont {J.-S.}\ \bibnamefont
  {Huang}}, \bibinfo {author} {\bibfnamefont {J.}~\bibnamefont {Kern}},
  \bibinfo {author} {\bibfnamefont {P.}~\bibnamefont {Geisler}}, \bibinfo
  {author} {\bibfnamefont {P.}~\bibnamefont {Weinmann}}, \bibinfo {author}
  {\bibfnamefont {M.}~\bibnamefont {Kamp}}, \bibinfo {author} {\bibfnamefont
  {A.}~\bibnamefont {Forchel}}, \bibinfo {author} {\bibfnamefont
  {P.}~\bibnamefont {Biagioni}}, \ and\ \bibinfo {author} {\bibfnamefont
  {B.}~\bibnamefont {Hecht.}},\ }\href@noop {} {\bibfield  {journal} {\bibinfo
  {journal} {Nano Lett.}\ }\textbf {\bibinfo {volume} {10}},\ \bibinfo {pages}
  {2105} (\bibinfo {year} {2010})}\BibitemShut {NoStop}%
\bibitem [{\citenamefont {Canfield}\ \emph {et~al.}(2007)\citenamefont
  {Canfield}, \citenamefont {Husu}, \citenamefont {Laukkanen}, \citenamefont
  {Bai}, \citenamefont {Kuittinen}, \citenamefont {Turunen},\ and\
  \citenamefont {Kauranen}}]{kauranen07}%
  \BibitemOpen
  \bibfield  {author} {\bibinfo {author} {\bibfnamefont {B.~K.}\ \bibnamefont
  {Canfield}}, \bibinfo {author} {\bibfnamefont {H.}~\bibnamefont {Husu}},
  \bibinfo {author} {\bibfnamefont {J.}~\bibnamefont {Laukkanen}}, \bibinfo
  {author} {\bibfnamefont {B.}~\bibnamefont {Bai}}, \bibinfo {author}
  {\bibfnamefont {M.}~\bibnamefont {Kuittinen}}, \bibinfo {author}
  {\bibfnamefont {J.}~\bibnamefont {Turunen}}, \ and\ \bibinfo {author}
  {\bibfnamefont {M.}~\bibnamefont {Kauranen}},\ }\href@noop {} {\bibfield
  {journal} {\bibinfo  {journal} {Nano Lett.}\ }\textbf {\bibinfo {volume}
  {7}},\ \bibinfo {pages} {1251} (\bibinfo {year} {2007})}\BibitemShut
  {NoStop}%
\bibitem [{\citenamefont {Czaplicki}\ \emph {et~al.}(2015)\citenamefont
  {Czaplicki}, \citenamefont {Mäkitalo}, \citenamefont {Siikanen},
  \citenamefont {Husu}, \citenamefont {Lehtolahti}, \citenamefont {Kuittinen},\
  and\ \citenamefont {Kauranen}}]{Czaplicki14}%
  \BibitemOpen
  \bibfield  {author} {\bibinfo {author} {\bibfnamefont {R.}~\bibnamefont
  {Czaplicki}}, \bibinfo {author} {\bibfnamefont {J.}~\bibnamefont
  {Mäkitalo}}, \bibinfo {author} {\bibfnamefont {R.}~\bibnamefont {Siikanen}},
  \bibinfo {author} {\bibfnamefont {H.}~\bibnamefont {Husu}}, \bibinfo {author}
  {\bibfnamefont {J.}~\bibnamefont {Lehtolahti}}, \bibinfo {author}
  {\bibfnamefont {M.}~\bibnamefont {Kuittinen}}, \ and\ \bibinfo {author}
  {\bibfnamefont {M.}~\bibnamefont {Kauranen}},\ }\href {\doibase
  10.1021/nl503901e} {\bibfield  {journal} {\bibinfo  {journal} {Nano Letters}\
  }\textbf {\bibinfo {volume} {15}},\ \bibinfo {pages} {530} (\bibinfo {year}
  {2015})}\BibitemShut {NoStop}%
\bibitem [{\citenamefont {de~Ceglia}\ \emph {et~al.}(2015)\citenamefont
  {de~Ceglia}, \citenamefont {Vincenti}, \citenamefont {Angelis}, \citenamefont
  {Locatelli}, \citenamefont {Haus},\ and\ \citenamefont
  {Scalora}}]{deCeglia:15}%
  \BibitemOpen
  \bibfield  {author} {\bibinfo {author} {\bibfnamefont {D.}~\bibnamefont
  {de~Ceglia}}, \bibinfo {author} {\bibfnamefont {M.~A.}\ \bibnamefont
  {Vincenti}}, \bibinfo {author} {\bibfnamefont {C.~D.}\ \bibnamefont
  {Angelis}}, \bibinfo {author} {\bibfnamefont {A.}~\bibnamefont {Locatelli}},
  \bibinfo {author} {\bibfnamefont {J.~W.}\ \bibnamefont {Haus}}, \ and\
  \bibinfo {author} {\bibfnamefont {M.}~\bibnamefont {Scalora}},\ }\href
  {\doibase 10.1364/OE.23.001715} {\bibfield  {journal} {\bibinfo  {journal}
  {Opt. Express}\ }\textbf {\bibinfo {volume} {23}},\ \bibinfo {pages} {1715}
  (\bibinfo {year} {2015})}\BibitemShut {NoStop}%
\bibitem [{\citenamefont {Hartmann}\ \emph {et~al.}(2013)\citenamefont
  {Hartmann}, \citenamefont {Piatkowski}, \citenamefont {Ciesielski},
  \citenamefont {Mackowski},\ and\ \citenamefont {Hartschuh}}]{Hartmann13}%
  \BibitemOpen
  \bibfield  {author} {\bibinfo {author} {\bibfnamefont {N.}~\bibnamefont
  {Hartmann}}, \bibinfo {author} {\bibfnamefont {D.}~\bibnamefont
  {Piatkowski}}, \bibinfo {author} {\bibfnamefont {R.}~\bibnamefont
  {Ciesielski}}, \bibinfo {author} {\bibfnamefont {S.}~\bibnamefont
  {Mackowski}}, \ and\ \bibinfo {author} {\bibfnamefont {A.}~\bibnamefont
  {Hartschuh}},\ }\href@noop {} {\bibfield  {journal} {\bibinfo  {journal} {ACS
  Nano}\ }\textbf {\bibinfo {volume} {7}},\ \bibinfo {pages} {10257} (\bibinfo
  {year} {2013})}\BibitemShut {NoStop}%
\bibitem [{\citenamefont {Dickson}\ and\ \citenamefont
  {Lyon}(2000)}]{dickson00}%
  \BibitemOpen
  \bibfield  {author} {\bibinfo {author} {\bibfnamefont {R.~M.}\ \bibnamefont
  {Dickson}}\ and\ \bibinfo {author} {\bibfnamefont {L.~A.}\ \bibnamefont
  {Lyon}},\ }\href@noop {} {\bibfield  {journal} {\bibinfo  {journal} {J. Phys.
  Chem. B}\ }\textbf {\bibinfo {volume} {104}},\ \bibinfo {pages} {6095}
  (\bibinfo {year} {2000})}\BibitemShut {NoStop}%
\bibitem [{\citenamefont {Song}\ \emph {et~al.}(2011)\citenamefont {Song},
  \citenamefont {Bouhelier}, \citenamefont {Bramant}, \citenamefont {Sharma},
  \citenamefont {Dujardin}, \citenamefont {Zhang},\ and\ \citenamefont {{Colas
  des Francs}}}]{SongACS11}%
  \BibitemOpen
  \bibfield  {author} {\bibinfo {author} {\bibfnamefont {M.}~\bibnamefont
  {Song}}, \bibinfo {author} {\bibfnamefont {A.}~\bibnamefont {Bouhelier}},
  \bibinfo {author} {\bibfnamefont {P.}~\bibnamefont {Bramant}}, \bibinfo
  {author} {\bibfnamefont {J.}~\bibnamefont {Sharma}}, \bibinfo {author}
  {\bibfnamefont {E.}~\bibnamefont {Dujardin}}, \bibinfo {author}
  {\bibfnamefont {D.}~\bibnamefont {Zhang}}, \ and\ \bibinfo {author}
  {\bibfnamefont {G.}~\bibnamefont {{Colas des Francs}}},\ }\href {\doibase
  10.1021/nn201648d} {\bibfield  {journal} {\bibinfo  {journal} {~{ACS} \
  {N}ano}\ }\textbf {\bibinfo {volume} {5}},\ \bibinfo {pages} {5874} (\bibinfo
  {year} {2011})}\BibitemShut {NoStop}%
\bibitem [{\citenamefont {Wei}\ and\ \citenamefont {Xu}(2012)}]{Xu12review}%
  \BibitemOpen
  \bibfield  {author} {\bibinfo {author} {\bibfnamefont {H.}~\bibnamefont
  {Wei}}\ and\ \bibinfo {author} {\bibfnamefont {H.}~\bibnamefont {Xu}},\
  }\href@noop {} {\bibfield  {journal} {\bibinfo  {journal} {Nanophot.}\
  }\textbf {\bibinfo {volume} {1}},\ \bibinfo {pages} {155} (\bibinfo {year}
  {2012})}\BibitemShut {NoStop}%
\bibitem [{\citenamefont {Taminiau}\ \emph {et~al.}(2011)\citenamefont
  {Taminiau}, \citenamefont {Stefani},\ and\ \citenamefont {van
  Hulst}}]{Taminiau11}%
  \BibitemOpen
  \bibfield  {author} {\bibinfo {author} {\bibfnamefont {T.~H.}\ \bibnamefont
  {Taminiau}}, \bibinfo {author} {\bibfnamefont {F.~D.}\ \bibnamefont
  {Stefani}}, \ and\ \bibinfo {author} {\bibfnamefont {N.~F.}\ \bibnamefont
  {van Hulst}},\ }\href {\doibase 10.1021/nl103828n} {\bibfield  {journal}
  {\bibinfo  {journal} {Nano Lett.}\ }\textbf {\bibinfo {volume} {11}},\
  \bibinfo {pages} {1020} (\bibinfo {year} {2011})}\BibitemShut {NoStop}%
\bibitem [{\citenamefont {Heinz}(1991)}]{HEINZ1991}%
  \BibitemOpen
  \bibfield  {author} {\bibinfo {author} {\bibfnamefont {T.~F.}\ \bibnamefont
  {Heinz}},\ }\href {\doibase
  http://dx.doi.org/10.1016/B978-0-444-88359-9.50011-9} {\emph {\bibinfo
  {title} {Nonlinear Surface Electromagnetic Phenomena}}},\ edited by\ \bibinfo
  {editor} {\bibfnamefont {H.-E.}\ \bibnamefont {Ponath}}\ and\ \bibinfo
  {editor} {\bibfnamefont {G.}~\bibnamefont {Stegeman}},\ \bibinfo {series}
  {Modern Problems in Condensed Matter Sciences}, Vol.~\bibinfo {volume} {29}\
  (\bibinfo  {publisher} {Elsevier},\ \bibinfo {year} {1991})\ pp.\ \bibinfo
  {pages} {353 -- 416}\BibitemShut {NoStop}%
\bibitem [{\citenamefont {Thyagarajan}\ \emph {et~al.}(2012)\citenamefont
  {Thyagarajan}, \citenamefont {Rivier}, \citenamefont {Lovera},\ and\
  \citenamefont {Martin}}]{Thyagarajan:12}%
  \BibitemOpen
  \bibfield  {author} {\bibinfo {author} {\bibfnamefont {K.}~\bibnamefont
  {Thyagarajan}}, \bibinfo {author} {\bibfnamefont {S.}~\bibnamefont {Rivier}},
  \bibinfo {author} {\bibfnamefont {A.}~\bibnamefont {Lovera}}, \ and\ \bibinfo
  {author} {\bibfnamefont {O.~J.}\ \bibnamefont {Martin}},\ }\href {\doibase
  10.1364/OE.20.012860} {\bibfield  {journal} {\bibinfo  {journal} {Opt.
  Express}\ }\textbf {\bibinfo {volume} {20}},\ \bibinfo {pages} {12860}
  (\bibinfo {year} {2012})}\BibitemShut {NoStop}%
\bibitem [{\citenamefont {Novotny}\ and\ \citenamefont
  {Hecht}(2006)}]{lukasbook}%
  \BibitemOpen
  \bibfield  {author} {\bibinfo {author} {\bibfnamefont {L.}~\bibnamefont
  {Novotny}}\ and\ \bibinfo {author} {\bibfnamefont {B.}~\bibnamefont
  {Hecht}},\ }\href@noop {} {\emph {\bibinfo {title} {Principles of
  Nano-Optics}}}\ (\bibinfo  {publisher} {Cambridge University Press},\
  \bibinfo {year} {2006})\BibitemShut {NoStop}%
\bibitem [{\citenamefont {Ginzburg}\ \emph
  {et~al.}(2015{\natexlab{b}})\citenamefont {Ginzburg}, \citenamefont
  {Krasavin}, \citenamefont {Wurtz},\ and\ \citenamefont
  {Zayats}}]{ginzburg15}%
  \BibitemOpen
  \bibfield  {author} {\bibinfo {author} {\bibfnamefont {P.}~\bibnamefont
  {Ginzburg}}, \bibinfo {author} {\bibfnamefont {A.~V.}\ \bibnamefont
  {Krasavin}}, \bibinfo {author} {\bibfnamefont {G.~A.}\ \bibnamefont {Wurtz}},
  \ and\ \bibinfo {author} {\bibfnamefont {A.~V.}\ \bibnamefont {Zayats}},\
  }\href {\doibase 10.1021/ph500362y} {\bibfield  {journal} {\bibinfo
  {journal} {ACS Photonics}\ }\textbf {\bibinfo {volume} {2}},\ \bibinfo
  {pages} {8} (\bibinfo {year} {2015}{\natexlab{b}})}\BibitemShut {NoStop}%
\bibitem [{\citenamefont {Marini}\ \emph {et~al.}(2013)\citenamefont {Marini},
  \citenamefont {Conforti}, \citenamefont {Valle}, \citenamefont {Lee},
  \citenamefont {Tran}, \citenamefont {Chang}, \citenamefont {Schmidt},
  \citenamefont {Longhi}, \citenamefont {Russell},\ and\ \citenamefont
  {Biancalana}}]{Marini:13}%
  \BibitemOpen
  \bibfield  {author} {\bibinfo {author} {\bibfnamefont {A.}~\bibnamefont
  {Marini}}, \bibinfo {author} {\bibfnamefont {M.}~\bibnamefont {Conforti}},
  \bibinfo {author} {\bibfnamefont {G.~D.}\ \bibnamefont {Valle}}, \bibinfo
  {author} {\bibfnamefont {H.~W.}\ \bibnamefont {Lee}}, \bibinfo {author}
  {\bibfnamefont {T.~X.}\ \bibnamefont {Tran}}, \bibinfo {author}
  {\bibfnamefont {W.}~\bibnamefont {Chang}}, \bibinfo {author} {\bibfnamefont
  {M.~A.}\ \bibnamefont {Schmidt}}, \bibinfo {author} {\bibfnamefont
  {S.}~\bibnamefont {Longhi}}, \bibinfo {author} {\bibfnamefont {P.~S.~J.}\
  \bibnamefont {Russell}}, \ and\ \bibinfo {author} {\bibfnamefont
  {F.}~\bibnamefont {Biancalana}},\ }\href
  {http://stacks.iop.org/1367-2630/15/i=1/a=013033} {\bibfield  {journal}
  {\bibinfo  {journal} {New J. Phys.}\ }\textbf {\bibinfo {volume} {15}},\
  \bibinfo {pages} {013033} (\bibinfo {year} {2013})}\BibitemShut {NoStop}%
\bibitem [{\citenamefont {Hassan}\ \emph {et~al.}(2013)\citenamefont {Hassan},
  \citenamefont {Bouhelier}, \citenamefont {Bernardin}, \citenamefont {{Colas
  des Francs}}, \citenamefont {Weeber}, \citenamefont {Dereux},\ and\
  \citenamefont {Espiau~de Lamaestre}}]{Hassan13}%
  \BibitemOpen
  \bibfield  {author} {\bibinfo {author} {\bibfnamefont {K.}~\bibnamefont
  {Hassan}}, \bibinfo {author} {\bibfnamefont {A.}~\bibnamefont {Bouhelier}},
  \bibinfo {author} {\bibfnamefont {T.}~\bibnamefont {Bernardin}}, \bibinfo
  {author} {\bibfnamefont {G.}~\bibnamefont {{Colas des Francs}}}, \bibinfo
  {author} {\bibfnamefont {J.-C.}\ \bibnamefont {Weeber}}, \bibinfo {author}
  {\bibfnamefont {A.}~\bibnamefont {Dereux}}, \ and\ \bibinfo {author}
  {\bibfnamefont {R.}~\bibnamefont {Espiau~de Lamaestre}},\ }\href {\doibase
  10.1103/PhysRevB.87.195428} {\bibfield  {journal} {\bibinfo  {journal} {Phys.
  Rev. B}\ }\textbf {\bibinfo {volume} {87}},\ \bibinfo {pages} {195428}
  (\bibinfo {year} {2013})}\BibitemShut {NoStop}%
\end{thebibliography}

%

\end{document}